\documentclass[amsmath,amssymb,aps,pre,twocolumn,superscriptaddress]{revtex4-1}

\newcommand{\lamL}{\lambda_{\text{$\ell$}}}
\newcommand{\lamH}{\lambda_{\text{h}}}
\newcommand{\xSL}{S_{\text{$\ell$}}}	
\newcommand{\xSH}{S_{\text{h}}}
\newcommand{\xIL}{\rho_{\text{$\ell$}}}
\newcommand{\xIH}{\rho_{\text{h}}}

\usepackage{graphicx}

\begin{document}

\title{Small inter-event times govern epidemic spreading on networks}

\author{Naoki Masuda}
\affiliation{Department of Mathematics
University at Buffalo, State University of New York, Buffalo, NY 14260-2900, USA}
\affiliation{Computational and Data-Enabled Science and Engineering Program,
University at Buffalo, State University of New York, Buffalo, NY 14260-5030, USA}
\email{naokimas@buffalo.edu}
\author{Petter Holme}
\affiliation{Tokyo Tech World Research Hub Initiative (WRHI), Institute of Innovative Research, Tokyo Institute of Technology, Yokohama 226-8503, Japan}

\begin{abstract}
Many aspects of human and animal interaction, such as the frequency of contacts of an individual, the number of interaction partners, and the time between the contacts of two individuals, are characterized by heavy-tailed distributions. These distributions affect the spreading of e.g.\ infectious diseases or rumors, often because of impacts of the right tail of the distributions (i.e.\ the large values). In this paper, we show that when it comes to inter-event time distributions, it is not the tail but the small values that control spreading dynamics.
We investigate this effect both analytically and numerically for different versions of the Susceptible-Infected-Recovered model on different types of networks.
\end{abstract}

\maketitle

\section{Introduction}

Despite the continuous advances in medicine, infectious diseases is still a major burden to the global health. It is, however, an area where physics modeling can be of tangible support to society.  Theoretical epidemiology has developed several core concepts that is guiding today's medical epidemiologists, such as: epidemic thresholds, herd immunity and the basic reproductive number~\cite{giesecke}. Temporal network epidemiology~\cite{Masuda2013F1000}---studying how structures in the time and network of contacts between people affect the spread of infectious diseases---is an emerging area that can help improving epidemic forecasting and intervention.

One salient feature of many empirical contact data sets is that the times of contacts cannot be described as a Poisson process~\cite{karsai2018bursty}. Specifically, the \textit{interevent times} (i.e., the time between contacts), both of pairs of individuals and of the individuals themselves, follow right-skewed fat-tailed distributions (distributions wider than the exponentially distributed ones of a Poisson process).
For inter-event times, the effect of a fat-tailed distribution can depend on details of the model---some papers find heterogeneous distributions facilitates spreading~\cite{Rocha2013PlosComputBiol,Vanmieghem2013PhysRevLett,JoPerotti2014PhysRevX,Masuda2018SiamRev,Mancastroppa2019JStatMech}, whereas others find the opposite effect~\cite{VazquezA2007PhysRevLett,Iribarren2009PhysRevLett,Karsai2011PhysRevE,Min2011PhysRevE,MinGohKim2013EPL,Horvath2014NewJPhys}. 
Typically, the effect of inter-event times is explained as an effect of the right tail of the inter-event times~\cite{VazquezA2007PhysRevLett,Iribarren2009PhysRevLett,Karsai2011PhysRevE,Rocha2013PlosComputBiol,Min2011PhysRevE,MinGohKim2013EPL,Miritello2011PhysRevE}. In this paper, however, we will show that the short-time end of the distribution impacts spreading more than the tail does.

Fat-tailed distributions can be approximated by a mixture of a small number of exponential distributions~\cite{Feldmann2002PerfEval,Jiang2016JStatMech,Okada2020RSocOpenSci}. We use this observation to develop an analytical framework to calculate epidemic thresholds for the SIR model when inter-event times obey a distribution whose variance can be tuned from Poissonian tails to infinitely fat ones.

\begin{figure}
\includegraphics[width=.8\columnwidth]{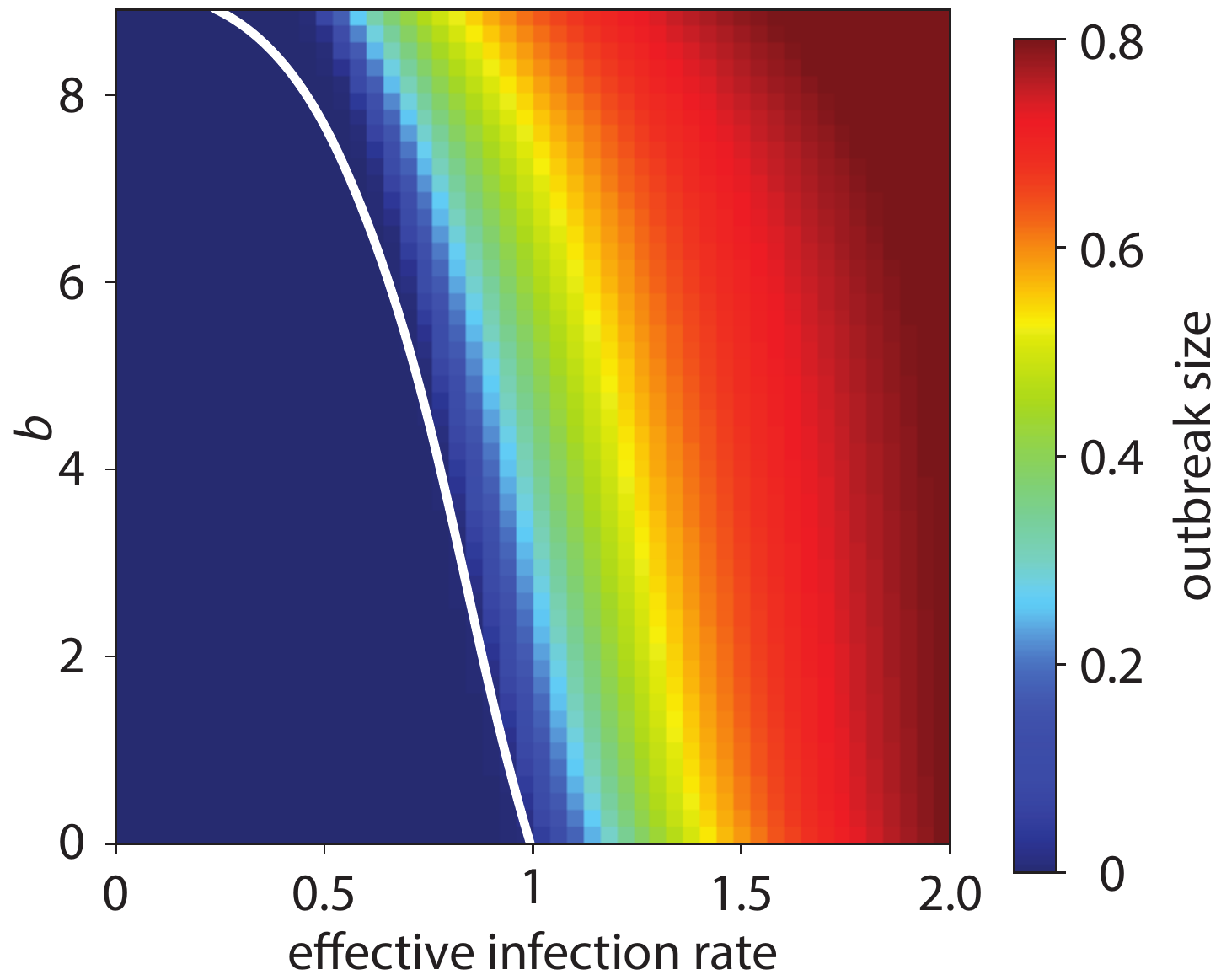}
\caption{Final outbreak size of the node-centric SIR model in the infinite well-mixed population with $a=0.1$, obtained by direct numerical solutions of the master equation. The solid line shows the analytically obtained epidemic threshold (Eq.~\eqref{eq:mu_c}).
}
\label{fig:theory}
\end{figure}

\section{Model}

We start by analyzing the following susceptible-infected-recovered (SIR) model with a general distribution of inter-event times attached on a static network with $N$ nodes. Each node is either susceptible, infected or recovered at any point of continuous time.
    Each node $i$ ($1\le i\le N$) carries an independent and identical point process whose probability density of inter-event times, denoted by $\tau$, is given by $\psi(\tau)$. When an event occurs at the $i$th node, the node is activated and selects one of its neighbors, denoted by $j$, uniformly randomly. The network is assumed to be undirected and unweighted. If either the $i$th or $j$th node is infected and the other is susceptible, the susceptible node becomes infected with probability $\beta$. Any infected node recovers according to a Poisson process with rate $\mu$. Once a node has recovered, it will not be reinfected or infect other nodes.
The present SIR model is node-centric in the sense that infection events are triggered by activation of either susceptible or infected nodes. 
In fact, inter-event times for single nodes, $\psi(\tau)$, obey fat-tailed distributions in various empirical data relevant to contagion, such as phone calls and email correspondences \cite{GohBarabasi2008EPL}. The present model approximates, for example, a situation in which an activated individual may call a randomly selected individual and the one with the newest information updates the other one. The same~\cite{JoPerotti2014PhysRevX,Masuda2018SiamRev,Mancastroppa2019JStatMech} or similar~\cite{Rocha2013PlosComputBiol} node-centric
infection mechanisms have been used in the literature. The well-studied ``activity driven model'' also assumes a node-centric mechanism, where one creates temporary undirected edges over which infection may occur in either direction~\cite{Perra2012SciRep,Zino2016PhysRevLett}. If $\psi(\tau)$ is exponential, each node process is Poissonian which reduces our model to the standard Markovian SIR model. In this case, the epidemic threshold in the well-mixed population is given by $2\beta/(\langle \tau\rangle \mu) = 1$, where the factor 2 results from the fact that each edge is selected for possible infection at a rate of $2/\langle \tau\rangle$ owing to the activation of either of the two nodes incident to the edge.

\begin{figure}
\includegraphics[width=\columnwidth]{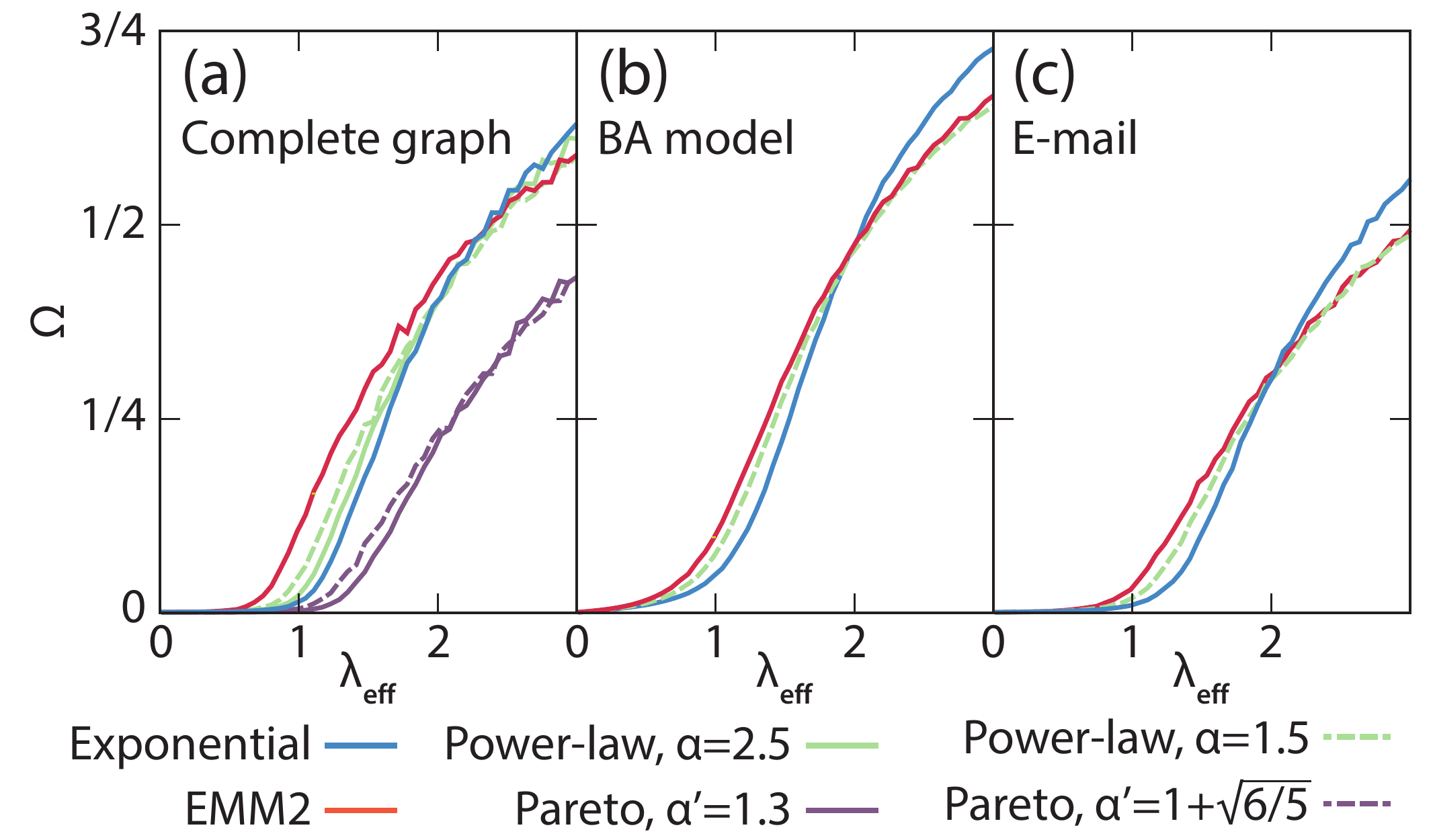}
\caption{Final outbreak size $\Omega$ of the node-centric SIR model on different networks as a function of the effective infection rate, $\lambda_{\rm eff}$.
(a) Complete graph with $N=10^3$ nodes.
(b) BA network having $N=10^3$ nodes, the number of edges that each added node introduces $m=3$, and the size of the initial clique $m_0 = 3$. (c) Largest connected component of an email network having
$N=1,133$ nodes and $5,451$ edges \cite{GuimeraDanon2003PhysRevE}. 
For PL1, we set $\alpha=1.5$ or $2.5$.
Assuming the equilibrium, we drew the distribution of waiting times, i.e.\ the time to the first activation from
$\psi^{\rm w}(\tau) = (\alpha-1)/(1+\tau)^{\alpha}$, independently for each node.
For the Pareto distribution in (a), we set $\alpha=1.3$ and $\alpha=1+\sqrt{6/5}$ and drew the time to the first activation from $\psi^{\rm w}(\tau) = (\alpha^{\prime}-1) \tau_0^{\alpha^{\prime}-1}/\tau^{\alpha{\prime}}$ ($\tau\ge \tau_0$) for each node; we also set $\tau_0 = 1$ without loss of generality,
because changing $\tau_0$ corresponds to rescaling the time. 
For EMM2, we set $a=0.3$, $\text{CV} = \sqrt{5}$, and drew the time to the first activation from
$\psi^{\rm w}(\tau) = \xSL^* \lamL \exp(-\lamL \tau) + \xSH^* \lamH \exp(-\lamH \tau)$.
For each of the six distributions of inter-event times, we scanned values of $\mu$ to span a range of $\lambda_{\rm eff}$. We measured $\Omega$ as the average over $n=5,000$ simulations. Error bars are omitted for clarity. They would be 3--5 times thicker than the lines.}
\label{fig:nR}
\end{figure}

\section{Results}
\subsection{Epidemic threshold for a mixture of exponential inter-event time distributions}

To analytically examine effects of a fat-tailed $\psi(\tau)$ on
the epidemic threshold, we assume an infinite well-mixed population and
represent fat-tailed distributions by a mixture of two exponential distributions with rates $\lamL$ and $\lamH (\ge \lamL)$. In other words, we set
\begin{equation}
\psi(\tau) = a \lamL \exp(-\lamL \tau) + (1-a) \lamH \exp(-\lamH \tau),
\label{eq:EMM k=2}
\end{equation}
where $0 < a < 1$ is the mixture weight. We refer to Eq.~\eqref{eq:EMM k=2} as a \textit{two-component exponential mixture model} (EMM2).
A large $\tau$ tends to be produced when the exponential distribution with the lower rate, $\lamL$, is selected with probability $a$, and vice versa. Across a scale of $\tau$, EMM2 mimics a fat-tailed $\psi(\tau)$ if $a$ is small and $\lamL \ll \lamH$.
Although EMM2 has, strictly speaking, a short tail,
it is often practically good at approximating empirical distributions of inter-event times~\cite{Feldmann2002PerfEval,Masuda2013Hawkes,Jiang2016JStatMech,Okada2020RSocOpenSci}.

We analyze the SIR dynamics combined with EMM2 inter-event times by assuming every node to take either of the two states, low or high, depending on which of $\lamL$ or $\lamH$ is currently employed for generating the next activation time.
Each node is assumed to transit between the two states. Upon its activation, each node interacts with a randomly selected different node, possibly transmitting the infection between them. Then, 
the activated node draws its new rate (i.e., either $\lamL$ or $\lamH$), hence the new state, according to which the time to the next activation of the node is generated. Because inter-event times are independently drawn from Eq.~\eqref{eq:EMM k=2} each time, the node upon the activation visits the low and high states with probability $a$ and $1-a$, respectively, regardless of the current state of the node.

We denote the fraction of the susceptible nodes in the low and high states among the $N$ nodes by $\xSL$ and $\xSH$, respectively, and the fraction of the infected nodes in the low and high states by $\xIL$ and $\xIH$, respectively. The fraction of the recovered nodes is equal to $1-\xSL-\xSH-\xIL-\xIH$.
The master equation representing the SIR dynamics is given by
\begin{widetext}
\begin{subequations}
\begin{align}
{\rm d}\xSL/{\rm d}t &=& - \left[\lamL \xSL (\xIL + \xIH) + (\lamL \xIL + \lamH \xIH) \xSL\right] \beta
+ \lamH \xSH \left[1- (\xIL + \xIH)\beta\right] a
- \lamL \xSL \left[1- (\xIL + \xIH)\beta\right] (1-a),
\label{eq:dxSL/dt}\\
{\rm d}\xSH/{\rm d}t &=& - \left[\lamH \xSH (\xIL + \xIH) + (\lamL \xIL + \lamH \xIH) \xSH\right] \beta
+ \lamL \xSL \left[1- (\xIL + \xIH)\beta\right] (1-a)
- \lamH \xSH \left[1- (\xIL + \xIH)\beta\right] a,
\label{eq:dxSH/dt}\\
{\rm d}\xIL/{\rm d}t  &=& (\lamL \xSL + \lamH \xSH)(\xIL + \xIH) \beta a
+ (\lamL \xIL + \lamH \xIH) \xSL \beta
+ \lamH \xIH a - \lamL \xIL (1-a) - \mu \xIL,
\label{eq:dxIL/dt}\\
{\rm d}\xIH/{\rm d}t &=& (\lamL \xSL + \lamH \xSH)(\xIL + \xIH) \beta (1-a)
+ (\lamL \xIL + \lamH \xIH) \xSH \beta
+ \lamL \xIL (1-a) - \lamH \xIH a - \mu \xIH,
\label{eq:dxIH/dt}
\end{align}
\end{subequations}
\end{widetext}
where $t$ denotes the time. The first term on the right-hand side of Eq.~\eqref{eq:dxSL/dt} corresponds to infection; $\lamL \xSL (\xIL + \xIH) \beta$ is the product of the rate $\lamL \xSL$ at which susceptible nodes in the low state is activated, the probability $\xIL + \xIH$ that
an infected node is selected as partner to interact with, and the probability $\beta$ that infection happens.
The rate $(\lamL \xIL + \lamH \xIH) \xSL \beta$ is the product of the rate $\lamL \xIL + \lamH \xIH$ at which infected nodes are activated, the probability $\xSL$ that a susceptible node is selected as partner, and the probability $\beta$ that infection happens.
The second term on the right-hand side of Eq.~\eqref{eq:dxSL/dt} represents the transition of a susceptible node from the high to the low state. The third term represents the transition of a susceptible node from the low to the high state.
Equations~\eqref{eq:dxSH/dt}, \eqref{eq:dxIL/dt}, and \eqref{eq:dxIH/dt} were similarly derived.

To derive the epidemic threshold, we consider the initial condition in which most nodes are susceptible and a small fraction of nodes is infected. The zeroth-order continuous-time Markov process representing the time course of each node's state has the transition rate from the low to high state of $\lamL (1-a)$ and that from the high to low state of $\lamH a$. Therefore, the stationary state of the two states before an infinitesimal fraction of infection is introduced to the population is given by
\begin{equation}
\xSL^* = \frac{\lamH a}{\lamL (1-a) + \lamH a}
\end{equation}
and
\begin{equation}
\xSH^* = \frac{\lamL (1-a)}
{\lamL (1-a) + \lamH a}.
\end{equation}
We perturb this stationary state by setting
$\xSL = \xSL^* + \Delta \xSL$, $\xSH = \xSH^* + \Delta \xSH$, $\xIL = \Delta \xIL$ and $\xIH = \Delta \xIH$, where $\Delta \xSL$, $\Delta \xSH$, $\Delta \xIL$, and $\Delta \xIH$ are small and satisfy
$\Delta \xSL + \Delta \xSH + \Delta \xIL + \Delta \xIH = 0$ due to the conservation of the mass at $t=0$. Denote by $J$ the Jacobian matrix
of the linearized dynamics around the disease-free initial condition.
If any of the eigenvalues of $J$ is positive, a macroscopic number of infected (and recovered) nodes will result, which is the condition that determines the epidemic threshold. As shown in Appendix~\ref{sec:threshold}, the condition under which a large-scale epidemic spreading can occur is given by
\begin{subequations}
\begin{equation}
\mu < \mu_{\rm c} \equiv \frac{\beta}{\langle \tau\rangle}
\left[1-\frac{1}{2\beta\gamma}
+ \sqrt{\left(1+\frac{1}{2\beta\gamma}\right)^2 + \frac{b}{\gamma}} \; \right],
\label{eq:mu_c}
\end{equation}
where $\mu_{\rm c}$ is the epidemic threshold in terms of the recovery rate,
\begin{equation}
\langle \tau \rangle = \frac{a}{\lamL} + \frac{1-a}{\lamH}
\label{eq:<tau>}
\end{equation}
is the mean inter-event time for each node, \begin{equation}
b = \frac{\langle \tau^2\rangle}{2\langle \tau\rangle^2} - 1,
\label{eq:b}
\end{equation}
and
\begin{equation}
\gamma = \left(1-\sqrt{\frac{ab}{1-a}}\right)
\left(1 + \sqrt{\frac{(1-a)b}{a}}\right).
\label{eq:def gamma}
\end{equation}
\end{subequations}

Equation~\eqref{eq:mu_c} indicates that, when $\lamL = \lamH$ such that $\psi(\tau)$ is an exponential distribution, one obtains $b=0$ such that the epidemic threshold in terms of the effective infection rate \cite{BogunaLafuerza2014PhysRevE}, i.e., $\lambda_{\rm eff} \equiv 2\beta/\langle\tau\rangle \mu$, is equal to 1. Crucially, Eq.~\eqref{eq:mu_c} also implies that $\mu_{\rm c} > 2\beta/\langle \tau\rangle$, or equivalently, $\lambda_{\rm eff} < 1$, if and only if $b > 0$. 
This is because
the two activation rates are expressed as 
\begin{subequations}
\begin{align}
\lamL = \langle \tau\rangle^{-1} \left(1 + \sqrt{\frac{(1-a)b}{a}}\right)^{-1}
\label{eq:lamL reparameterized}
\end{align}
and
\begin{align}
\lamH = \langle \tau\rangle^{-1} \left(1 - \sqrt{\frac{ab}{1-a}}\right)^{-1},
\label{eq:lamH reparameterized}
\end{align}
\end{subequations}
and $\lamH > 0$ guarantees $\gamma > 0$.
Because $b$ represents the extent to which $\tau$ is heterogeneously distributed (related to the coefficient of variation, CV, via $b=(\text{CV}^2-1)/2$), these results mean that infection  spread more easily for heterogeneously distributed than exponentially distributed inter-event times, within the framework of EMMs.

\subsection{Numerical validation of the analytical results}

To validate the theory, we numerically integrated the master equations (Eqs.~\eqref{eq:dxSL/dt}--\eqref{eq:dxIH/dt}) with the initial conditions
\begin{subequations}
\begin{align}
\xSL &=& (1-\rho_0) \lamH a/\left[\lamL (1-a) + \lamH a\right],\\
    \xSH &=& (1-\rho_0)\lamL (1-a)/\left[\lamL (1-a) + \lamH a\right],\\
    \xIL &=& \rho_0 \lamH a/\left[\lamL (1-a) + \lamH a\right],\\
    \xIH &=& \rho_0 \lamL (1-a)/\left[\lamL (1-a) + \lamH a\right],
\end{align}
\end{subequations}
where $\rho_0=10^{-3}$ is the initial fraction of infected nodes.
For the subsequent analysis, we set $a=0.1$ and $\beta=1$.
The final fraction of recovered nodes, also known as the final outbreak size, denoted by $\Omega$, is estimated as the fraction of recovered nodes at a sufficiently large time, $t=10^3$. In Fig.~\ref{fig:theory}, we show $\Omega$ as a function of the effective infection rate $2(\langle \tau\rangle \mu)^{-1}$ and $b$. Note that, for EMM2 it holds that
$b \ge 0$ \cite{Yannaros1994AnnInstStatMath} and $b < (1-a)/a = 9$.

Figure~\ref{fig:theory} confirms the theoretically derived epidemic threshold (solid line) and indicates that $\Omega$ increases with $b$. This result is consistent with the previous numerical results obtained with the same~\cite{Masuda2018SiamRev} or similar~\cite{Rocha2013PlosComputBiol} node-centric SIR models, the theoretical results derived from branching processes under the assumption of tree-like networks
\cite{JoPerotti2014PhysRevX}, and the theoretical results derived for the susceptible-infected-susceptible (SIS) model for a variant of the activity-driven network model of temporal networks~\cite{Mancastroppa2019JStatMech}.

In the limit $b\to (1-a)/a$, which corresponds to the largest possible dispersion in inter-event times given by Eq.~\eqref{eq:b}, one obtains $\gamma\to 0$. In this limit, one obtains
$\lamL \to a/\langle \tau\rangle$ and $\lamH \to\infty$, and
Eq.~\eqref{eq:mu_c} suggests that $\mu_{\rm c}$ approaches $(2\beta/\langle \tau\rangle)\times (1+b\beta/2)$, where $b = (1-a)/a$. Therefore, the epidemic threshold in terms of $\mu$ is 
$(1+b\beta/2)$ times larger than that for the Poissonian SIR dynamics, which corresponds to $b=0$. In particular, when $a\to 0$, one obtains $b\to\infty$, such that $\mu_{\rm c}$ diverges. For such a distribution of inter-event times, a large-scale epidemic spreading can occur even when $\lambda_{\rm eff}$ is tiny. This phenomenon is reminiscent of the vanishing epidemic threshold in the Poissonian SIS and SIR dynamics in scale-free networks \cite{Pastorsatorras2015RevModPhys}. However, the mechanism is different. In the present model, the epidemic threshold vanishes because infinitesimally short inter-event times are dominant, not because a fat-tailed distribution occasionally produces large values as in the case of epidemic spreading in static scale-free networks \cite{Pastorsatorras2015RevModPhys}. A similar result was found in Ref.~\cite{JoPerotti2014PhysRevX}. Note that a fat-tailed distribution is not needed for reaching this conclusion; EMM2 does not have a fat tail.

\subsection{Numerical validation of importance of short inter-event times}

To further establish the importance of the short-$\tau$ end of the distribution, we carried out simulations of the SIR dynamics on the complete graph with $N=10^3$ nodes with different power-law $\psi(\tau)$ having different lower bounds.
We started each simulation from the initial condition in which one randomly selected node is infected and all the other nodes are susceptible. We drew the time to the first activation of each node from the waiting-time distribution corresponding to $\psi(\tau)$
\cite{Takaguchi2011PhysRevE,JoPerotti2014PhysRevX}.

Consider a power-law distribution of inter-event times given by
\begin{equation}
\psi(\tau) = \frac{\alpha}{(1+\tau)^{\alpha+1}},
\end{equation}
which we denote by PL1.
PL1 has support $\tau\ge 0$ and therefore has some probability mass near $\tau=0$, being able to produce short inter-event times with some probability.
With $\alpha=2.5$, PL1 has $b=1/(\alpha-2)=2/3$ and $\text{CV} = \sqrt{\alpha/(\alpha-2)} = \sqrt{5}$. 
The Pareto distribution is another popular power-law distribution. It is given by
\begin{equation}
\psi(\tau) =
\begin{cases}
\alpha^{\prime} \tau_0^{\alpha^{\prime}} / \tau^{\alpha^{\prime}+1} & (\tau \ge \tau_0),\\
0 & (\tau < \tau_0),
\end{cases}
\end{equation}
and is not capable of producing inter-event times shorter than $\tau_0$.
For this distribution, one obtains $b=(\alpha^{\prime}-1)^2/\left[2\alpha^{\prime}(\alpha^{\prime}-2)\right]$ and
$\text{CV} = \left[ \alpha^{\prime}(\alpha^{\prime}-2)\right]^{-1/2}$. The $b$ value of $2/3$ is produced by $\alpha^{\prime} = 1+\sqrt{6/5} \approx 2.095$. Thus, we can compare two power-law distributions sharing the $b$ (and CV) value but are considerably different at both small and large $\tau$.

In Fig.~\ref{fig:nR}, we show $\Omega$ for a range of $\lambda_{\rm eff}$ for the two power-law $\psi(\tau)$ and the exponential $\psi(\tau)$ with equal mean $\tau$. The figure shows that that PL1 enhances epidemics as compared to the Poisson case, at least near the epidemic threshold, whereas the Pareto distribution suppresses epidemics. We believe that this is because the Pareto distribution is not capable of generating short inter-event times, whereas PL1 is.
Note that the present Pareto distribution has a longer tail than PL1, i.e., $\alpha^{\prime} < \alpha$. Therefore, the capability for a $\psi(\tau)$ to generate short inter-event times is not a consequence of having a fat tail. To further support this claim, we also simulated the case in which
$\psi(\tau)$ is EMM2 with
$a=0.3$ and $\text{CV} = \sqrt{5}$, which automatically determines $\lamL$ and $\lamH$ via
Eqs.~\eqref{eq:lamL reparameterized} and \eqref{eq:lamH reparameterized}. The outbreak size with this $\psi(\tau)$ is shown in Fig.~\ref{fig:nR}.
We find that  $\Omega$ is much larger than in the case of  power-law distributions with the same mean across the entire range of $\lambda_{\rm eff}$. We emphasize that EMM2 does not have a fat tail in a rigorous sense. Qualitatively, the same results hold true even when $\langle \tau^2\rangle$ diverges.
Results for PL1 with $\alpha=1.5$ and those for the Pareto distribution with $\alpha^{\prime} = 1.3$ are shown by the dotted lines in Fig.~\ref{fig:nR}.

To summarize these results, although we expressed our theoretical results in terms of $b$, a larger value of $b$ or a fat tail of $\psi(\tau)$ is not a key factor impacting the epidemic threshold or the outbreak size. It is the size and the probability mass of $\tau$ at small $\tau$ values that play the key role.

We also confirmed qualitatively similar results for networks with a fat-tail degree distribution generated by the Barab\'{a}si-Albert (BA) model \cite{Barabasi1999Science} and the largest connected component of an email social network
\cite{GuimeraDanon2003PhysRevE} (Figs.~\ref{fig:nR}(b) and \ref{fig:nR}(c)).

\subsection{SIS model}

To complement the above results for the SIR model, we also numerically investigated the node-centric SIS model. As in the node-centric SIR model,
when an event occurs at the $i$th node, the node is activated and selects one of its neighbors, denoted by $j$, uniformly randomly. If either the $i$th or $j$th node is infected and the other is susceptible, the susceptible node contracts infection with probability $\beta$, which we set to $1$ for simplicity. Each infected node recovers at rate $\mu$ to become susceptible.

We assumed that all the nodes were initially infected in each simulation. We ran the simulation until either $t=t_{\max} = 300$ was reached or all the nodes became susceptible. In this set of simulations, we used the forms of $\psi(\tau)$ such that the mean inter-event time, $\langle \tau \rangle$, depended on the type of distribution (i.e., exponential, PL1, Pareto, or EMM2), and we varied $\mu$ to set values of $\lambda_{\rm eff}$. Therefore, values of the time do not bear a meaning. However, we confirmed that the results barely changed with $t_{\max}=1000$.
At $t=t_{\max}$, we measured the fraction of infected nodes. The prevalence, i.e., equilibrium fraction of infected nodes, is given by the fraction of infected nodes averaged over $n=500$ simulations.

\begin{figure}
\includegraphics[width=\columnwidth]{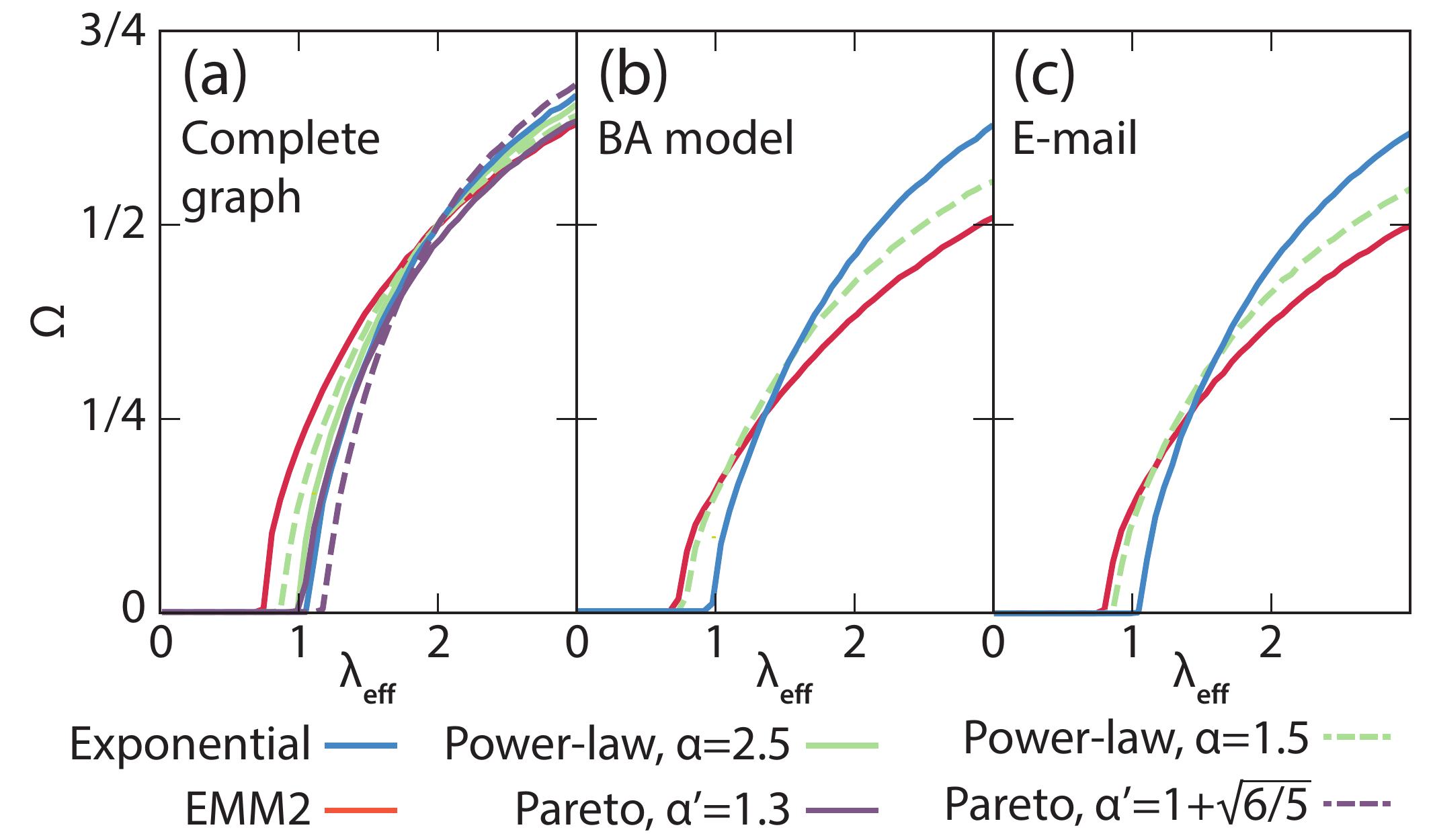}
\caption{Prevalence of the node-centric SIS model on different networks as a function of the effective infection rate, $\lambda_{\rm eff}$.
(a) Complete graph with $N=10^3$ nodes.
(b) BA network having $N=10^3$ nodes, the number of edges that each added node introduces $m=3$, and the size of the initial clique $m_0 = 3$. (c) Largest connected component of an email network having
$N=1,133$ nodes and $5,451$ edges \cite{GuimeraDanon2003PhysRevE}. 
For EMM2, we set $a=0.3$ and $\text{CV} = \sqrt{5}$.
Error bars are omitted for clarity. They would be 3--5 times thicker than the lines.}
\label{fig:SIS}
\end{figure}

The prevalence for the three networks is shown in Fig.~\ref{fig:SIS}. The results are similar to those for the SIR model (Fig.~\ref{fig:nR}). A notable difference to the case of the SIR model is that, for the complete graph, the Pareto distribution with $\alpha^{\prime}=1.3$ produces a slightly smaller epidemic threshold than the exponential distribution (Fig.~\ref{fig:SIS}(a)). Otherwise, these results support our main conclusion that the epidemic threshold is lessened by the capability of $\psi(\tau)$ to produce a larger fraction of short inter-event times than the exponential distribution with the same mean.
The results are also consistent with a previous study employing the activity-driven model of temporal networks in that burstiness of inter-event times decreases the epidemic threshold, increases the prevalence at small $\lambda_{\rm eff}$, and decreases the prevalence at large $\lambda_{\rm eff}$ \cite{Mancastroppa2019JStatMech}.

\subsection{Numerical results for edge-centric SIR models\label{sub:SIR}}

Many previous studies of the effect of inter-event times on epidemics found that fat-tailed distributions slow down the dynamics~\cite{VazquezA2007PhysRevLett,Iribarren2009PhysRevLett,Karsai2011PhysRevE,MinGohKim2013EPL,Horvath2014NewJPhys}. These studies all used \textit{edge-centric} models, where the next contact is generated for individual edges rather than nodes. In edge-centric models, delay or reduction in epidemic spreading is attributed to the waiting-time paradox, with which a newly infected node has to wait for long time (i.e., averaging waiting time larger than the mean inter-event time, $\langle \tau\rangle$) Below, we show that also in this situation it is the short-time end of the distribution of inter-event times that controls the dynamics. With extreme distributions having a large probability mass near 0, the outbreak size for a standard edge-centric SIR model can be even magnified compared to the Poissonian case with the same mean $\langle \tau\rangle$.

\begin{figure}
 \includegraphics[width=8cm]{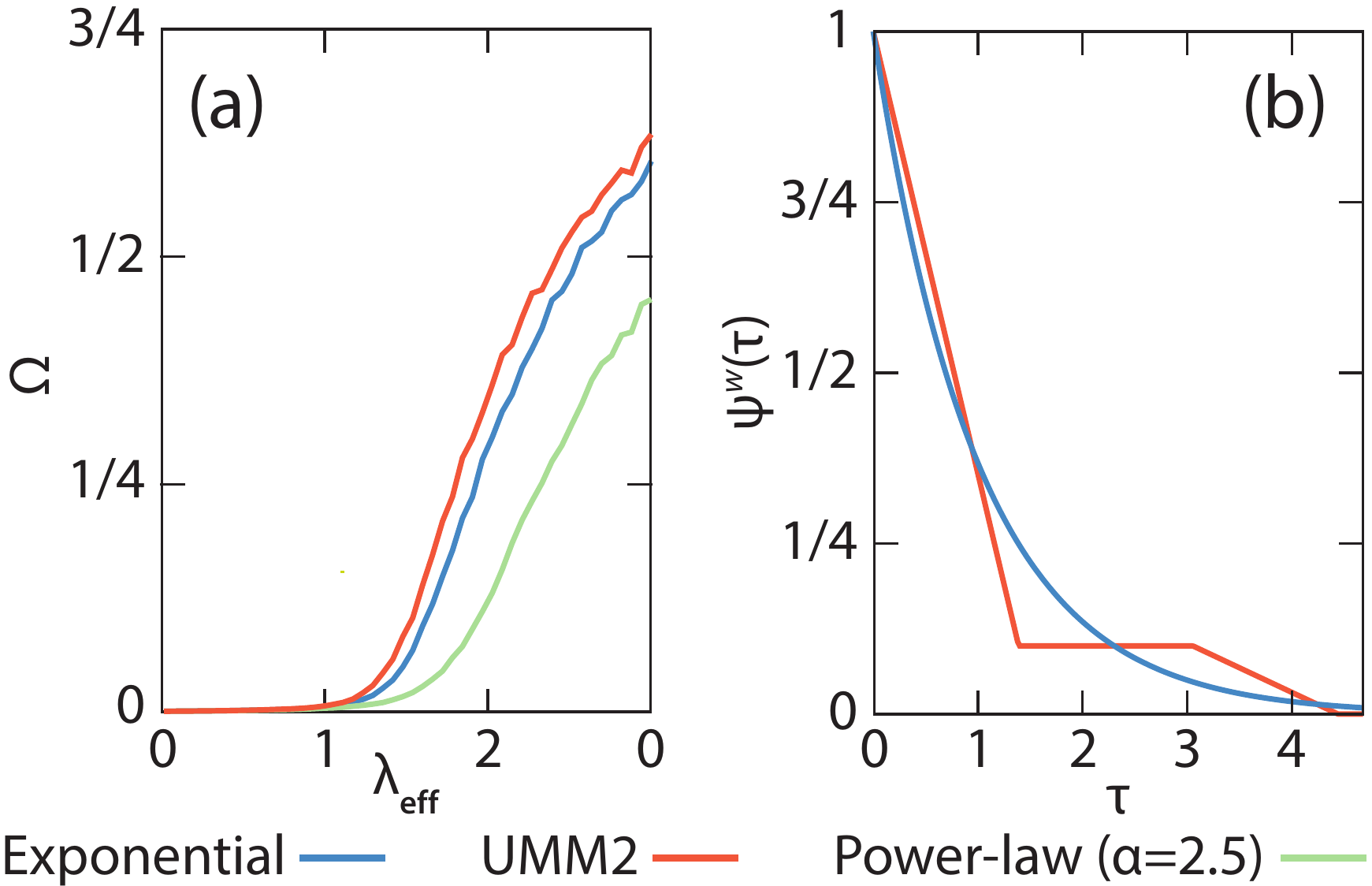}
\caption{Edge-centric SIR model. (a) Final size. We assumed a regular random graph with $N = 10^3$ nodes and mean degree five. (b) Waiting-time distribution,  $\psi^{\rm w}(\tau)$, for the exponential
distribution and the UMM2, both having $\langle\tau\rangle = 1$ and $\langle\tau^2\rangle = 2$. For the UMM2, we set $a = 1$, which yields $\epsilon\approx 1.389$.}
\label{fig:edge-centric}
\end{figure}

In edge-centric epidemic processes, if a node $i$ infects its neighbor $j$, the time $t$ to the next infection event on edge $(j, j')$, where $j' \neq i$, obeys the waiting-time distribution given by
\begin{equation}
\psi^{\rm w}(\tau) =
\frac{1}{{\langle\tau\rangle}} \int_{\tau}^\infty \psi(\tau^{\prime}) {\rm d}\tau^{\prime}
\label{eq:waiting-time distribution}
\end{equation}
rather than $\psi(\tau)$ \cite{Feller1971book2,Allen1990book,Masuda2016book}. 
The mean waiting time is given by
$\langle\tau^2\rangle / (2\langle\tau\rangle)$, where $\langle \cdot \rangle$ is the expectation under $\psi(\tau)$, and it is always larger than the half of the mean inter-event time, $\langle \tau\rangle / 2$, unless $\psi(\tau)$ is the delta distribution.
In particular, when $\psi(\tau)$ is fat-tailed, the mean waiting time is much larger than $\langle \tau\rangle / 2$ because $\langle\tau^2\rangle \gg \langle \tau\rangle^2$. Then, typical attempts to relay infection to a new node would take much longer time than $\langle \tau\rangle$.
This is the reason why fat-tailed $\psi(\tau)$ suppresses epidemic spreading in edge-centric epidemic processes.

In contrast, the node-centric SIR model circumvents the waiting-time paradox. If a susceptible node $v_i$ is activated and infected from $v_j$, then $v_i$ draws a time to the next activation from distribution $\psi(\tau)$, not from $\psi^{\rm w}(\tau)$, potentially infecting $v_i$'s other neighbors. In this manner, infection can spread rapidly in arbitrary networks.

We compare the final size for the edge-centric SIR model when 
$\psi(\tau)$ is exponential or PL1 with $\alpha=2.5$, sharing the mean, $\langle \tau\rangle=1$, in  Fig.~\ref{fig:edge-centric}(a). Confirming the results of the previous studies, the final size is substantially smaller in the case of PL1 than the exponential distribution. Given our insights we have obtained for the node-centric SIR model, is it possible to enhance epidemic spreading in the edge-centric SIR model by devising $\psi(\tau)$?

A waiting-time distribution $\psi^{\rm w}(\tau)$ may have a certain amount of probability mass at small $\tau$ depending on $\psi(\tau)$. We consider that in conventional edge-centric SIR models, the effect of the waiting-time paradox outweighs that of the probability mass of waiting times at small $\tau$ such that epidemic spreading would be suppressed by a long-tailed $\psi(\tau)$. Therefore, we considered a $\psi(\tau)$ which is little influenced by the waiting-time paradox while it has a relatively high probability mass at small waiting times. We created such a distribution as a mixture of two uniform densities, which we call the UMM2 (UMM for uniform mixture model). We define the UMM2 by
\begin{equation}
\psi(\tau) =
\begin{cases}
\frac{1-a}{\epsilon} & 0 \le \tau < \epsilon,\\
\frac{a}{\epsilon} & \frac{2-\epsilon}{2a} \le \tau < \frac{2-\epsilon}{2a} + \epsilon,\\
0 & \text{otherwise}.
\end{cases}
\end{equation}
Here we use UMM2 rather than EMM2 to suppress the effect of the waiting-time paradox; unless $\epsilon$ is large, $\langle \tau^2\rangle$ is smaller for a UMM2 than an EMM2 given the same mean $\langle \tau \rangle$. Then, we imposed $\langle \tau^2 \rangle = 2$, such that $b=0$ and $\text{CV} = 1$ as in the exponential distribution. The calculations shown in Appendix~\ref{sec:UMM2 analysis} gives
\begin{equation}
\epsilon = \frac{2\left[3(1-a)-\sqrt{3a(2-a)}\right]} {3-2a}.
\end{equation}
In this way, one can compare the UMM2 and the exponential distribution that share the mean inter-event time $\langle \tau \rangle = 1$ and the mean waiting time $\langle \tau^2\rangle/(2\langle \tau\rangle)= 1$, while the two distributions have different probability masses at small waiting times.

For the exponential distribution, one obtains $\psi(\tau) = \psi^{\rm w}(\tau) = \exp(-\tau)$. When $\psi(\tau)$ is the UMM2, one obtains
\begin{equation}
\psi^{\rm w}(\tau) =
\begin{cases}
- \frac{1-a}{\epsilon}\tau + 1 & 0 \le \tau < \epsilon,\\
a & \epsilon \le \tau < \frac{2-\epsilon}{2a},\\
-\frac{a}{\epsilon}\tau + a + \frac{2-\epsilon}{2\epsilon} & \frac{2-\epsilon}{2a} \le \tau < \frac{2-\epsilon}{2a} + \epsilon,\\
0 & \frac{2-\epsilon}{2a}+\epsilon \le \tau.
\end{cases}
\end{equation}
Because $\psi^{\rm w}(0) = 1$ for both exponential distribution and UMM2, by letting $\epsilon$ large, one can create a relatively large probability mass of the waiting-time for the UMM2 at small $\tau$ values as compared to the exponential distribution (Fig.~\ref{fig:edge-centric}(b)).

We set $a=0.1$, which yields $\epsilon \approx 1.389$. The final size for this UMM2 is shown by the red line in Fig.~\ref{fig:edge-centric}(a). The figure suggests that the final size is larger for this distribution as compared to the exponential distribution. We note that the two distributions have the same $\langle \tau\rangle$ and $\langle \tau^2\rangle$.

\subsection{Epidemic threshold for two variants of node-centric SIR models}

The analysis of the edge-centric SIR model shown in Section~\ref{sub:SIR} supports the generality of the proposed mechanism to enhance epidemic spreading with non-exponential distributions of inter-event times, beyond a particular type of the node-centric SIR model. To further confirm this finding, we also analyzed two variants of the node-centric SIR model (Appendix~\ref{sec:variants}). Each of these models is partially subjected to the effect of the waiting-time paradox, but not as strongly so as the edge-centric SIR model analyzed above. For both variants of the node-centric SIR model, we have found that the epidemic threshold for the well-mixed population when $\psi(\tau)$ is EMM2 is independent of the form of $\psi(\tau)$ and equal to that for the exponential $\psi(\tau)$ with the same $\langle \tau\rangle$.

\section{Conclusions}

The literature about how fat-tailed inter-event time distributions affect epidemic spreading on networks has been inconclusive. In the present study, we have resolved this issue by attributing the ease of spreading to the balance between the waiting-time paradox and the abundance of short inter-event times.  Our results account for both cases where infection events are triggered by node activation (i.e., node-centric) or edge activation (i.e., edge-centric). In particular, in the node-centric SIR model (that is only weakly influenced by the waiting-time paradox), we have found that it is the frequency of short inter-event times, not a fat tail of $\psi(\tau)$, that determines the epidemic threshold and $\Omega$. We have reached these conclusions via a novel technique using a mixture of exponential distributions, capable of mimicking fat-tailed distributions. We expect that the same technique is applicable for analyzing other dynamical processes such as stochastic opinion dynamics, complex contagions, and coevolutionary dynamics in empirical and model temporal networks. This approach also lends itself well to simulation studies as the dynamics for well-mixed populations is exactly solvable by integrating the master equations. One can achieve even more realism with the use of exponential mixture models with more than two components  \cite{Feldmann2002PerfEval,Okada2020RSocOpenSci}.

\begin{acknowledgments}
P.H. was supported by JSPS KAKENHI Grant Number JP 18H01655 and by the Grant for Basic Science Research Projects by the Sumitomo Foundation.
\end{acknowledgments}

\appendix

\section{Derivation of the epidemic threshold}
\label{sec:threshold}

\begin{widetext}
The SIR dynamics linearized around the disease-free steady state is given by
\begin{equation}
\frac{{\rm d}}{{\rm d}t}
\begin{pmatrix}
\Delta \xSL\\ \Delta \xSH\\ \Delta \xIL\\ \Delta \xIH
\end{pmatrix}
=
J \begin{pmatrix}
\Delta \xSL\\ \Delta \xSH\\ \Delta \xIL\\ \Delta \xIH
\end{pmatrix}
\equiv
\begin{pmatrix}
- \lamL (1-a) & \lamH a & -2\lamL \xSL^* \beta & -(\lamL + \lamH) \xSL^* \beta\\
\lamL (1-a) & - \lamH a & - (\lamL + \lamH) \xSH^* \beta & -2\lamH \xSH^* \beta\\
0 & 0 & J_{33} & J_{34}\\
0 & 0 & J_{43} & J_{44}
\end{pmatrix}
\begin{pmatrix}
\Delta \xSL\\ \Delta \xSH\\ \Delta \xIL\\ \Delta \xIH
\end{pmatrix},
\end{equation}
where $J$ is the Jacobian matrix, and
\begin{subequations}
\begin{align}
J_{33} =& (\lamL \xSL^* + \lamH \xSH^*) \beta a + \lamL \xSL^* \beta - \lamL (1-a) - \mu,\\
J_{34} =& (\lamL \xSL^* + \lamH \xSH^*) \beta a + \lamH \xSL^* \beta + \lamH a,\\
J_{43} =& (\lamL \xSL^* + \lamH \xSH^*) \beta (1-a) + \lamL \xSH^* \beta + \lamL (1-a),\\
J_{44} =& (\lamL \xSL^* + \lamH \xSH^*) \beta (1-a) + \lamH \xSH^* \beta - \lamH a - \mu.
\end{align}
\end{subequations}

The leading principal minor of $J$ of order $2$ gives two eigenvalues of $J$, which are equal to 0 and
$-\left[ \lamL(1-a) + \lamH a \right] < 0$. The zero eigenvalue reflects the constraint 
$\Delta \xSL + \Delta \xSH + \Delta \xIL + \Delta \xIH = 0$.

The eigenvalues from the lower $2\times 2$ diagonal of $J$ are equal to 
$-\mu+ \Lambda_1$ and $-\mu+\Lambda_2$, where $\Lambda_1$ and $\Lambda_2 (> \Lambda_1)$ are the solutions of
\begin{equation}
\Lambda^2 + \left[ \lamL (1-a) + \lamH a - \frac{2\lamL \lamH \beta}{\lamL (1-a) + \lamH a} \right]\Lambda
- 2 \lamL \lamH \beta - \frac{\lamL \lamH(\lamH - \lamL)^2 a(1-a) \beta^2}
{\left[\lamL (1-a) + \lamH a\right]^2} = 0. 
\label{eq:Lambda}
\end{equation}
Therefore, the condition under which the disease-free steady state is destabilized is given by $-\mu + \Lambda_2 > 0$, which gives 
Eq.~\eqref{eq:mu_c}.
\end{widetext}

\section{Epidemic threshold and final size for the two variants of the node-centric SIR model\label{sec:variants}}

Assume that the inter-event time of each node independently obeys $\psi(\tau)$, that a node $v_i$ is activated, and that one of its neighbor, denoted by $v_j$, is uniformly randomly selected. In the original version of the node-centric SIR model, if either $v_i$ or $v_j$ is infected but not both, the infection between $v_i$ and $v_j$ occurs with probability $\beta$.
In the first new variant we consider, an infection happens with probability $\beta$ if and only if
$v_i$ is susceptible and $v_j$ is infected. In the second variant, infection happens with probability $\beta$ if and only if $v_i$ is infected and $v_j$ is susceptible. 

These node-centric SIR models are partially affected by the waiting-time paradox in different manners. In the first variant, if $v_i$ is infected by $v_j$, the time until $v_i$ contacts its neighbor $v_{\ell}$ for possibly infecting $v_{\ell}$ (if $v_{\ell}$ is susceptible) obeys the waiting-time distribution. Therefore, possible infection of $v_{\ell}$ by $v_i$ tends to be delayed, such that epidemic spreading may be suppressed by a fat-tailed $\psi(\tau)$. In contrast, epidemic spreading may be enhanced in this model because, if $v_i$ does not get infected by $v_j$, then $v_i$ may draw a short inter-event time according to $\psi(\tau)$ to be infected by its neighbor. In the second variant, if $v_i$ infects $v_j$, then the time to the next activation of $v_j$, which may let $v_j$ to infect another node, obeys the waiting-time distribution. Therefore, this secondary infection process may happen late. In contrast, epidemic spreading in the same model may be enhanced because, after infecting $v_j$, node $v_i$ draws the time to its next activation according to $\psi(\tau)$ to possibly infect its different neighbor.

Consider the first variant of the node-centric SIR model. When the population is infinite and well-mixed and $\psi(\tau)$ is an EMM2, the stochastic dynamics of epidemic spreading are given by
\begin{subequations}
\begin{align}
\frac{{\rm d}\xSL}{{\rm d}t} =& - \lamL \xSL (\xIL + \xIH) \beta
+ \lamH \xSH \left[1- (\xIL + \xIH)\beta\right] a\notag\\
&- \lamL \xSL \left[1- (\xIL + \xIH)\beta\right] (1-a),
\label{eq:dxSL/dt VM}\\
\frac{{\rm d}\xSH}{{\rm d}t} =& - \lamH \xSH (\xIL + \xIH) \beta
+ \lamL \xSL \left[1- (\xIL + \xIH)\beta\right] (1-a)\notag\\
&- \lamH \xSH \left[1- (\xIL + \xIH)\beta\right] a,
\label{eq:dxSH/dt VM}\\
\frac{{\rm d}\xIL}{{\rm d}t} =& (\lamL \xSL + \lamH \xSH)(\xIL + \xIH) \beta a
+ \lamH \xIH a\notag\\
&- \lamL \xIL (1-a) - \mu \xIL,
\label{eq:dxIL/dt VM}\\
\frac{{\rm d}\xIH}{{\rm d}t} =& (\lamL \xSL + \lamH \xSH)(\xIL + \xIH) \beta (1-a)
+ \lamL \xIL (1-a)\notag\\
&- \lamH \xIH a - \mu \xIH.
\label{eq:dxIH/dt VM}
\end{align}
\end{subequations}

Similarly, the stochastic dynamics of the second variant of the node-centric SIR model are given by
\begin{subequations}
\begin{align}
\frac{{\rm d}\xSL}{{\rm d}t} =& - (\lamL \xIL + \lamH \xIH) \xSL \beta
+ \lamH \xSH a - \lamL \xSL (1-a),
\label{eq:dxSL/dt IP}\\
\frac{{\rm d}\xSH}{{\rm d}t} =& - (\lamL \xIL + \lamH \xIH) \xSH \beta
+ \lamL \xSL (1-a) - \lamH \xSH a,
\label{eq:dxSH/dt IP}\\
\frac{{\rm d}\xIL}{{\rm d}t} =& (\lamL \xIL + \lamH \xIH) \xSL \beta
+ \lamH \xIH a - \lamL \xIL (1-a) - \mu \xIL,
\label{eq:dxIL/dt IP}\\
\frac{{\rm d}\xIH}{{\rm d}t} =& (\lamL \xIL + \lamH \xIH) \xSH \beta
+ \lamL \xIL (1-a) - \lamH \xIH a - \mu \xIH.
\label{eq:dxIH/dt IP}
\end{align}
\end{subequations}

The eigenvalues of the Jacobians of these systems evaluated at the disease-free equilibrium determine the epidemic threshold. For both SIR models, the eigenequation to determine the two eigenvalues resulting from the leading principal minor of size two is the same that of the original SIR model. In addition, for both variants, the other two eigenvalues of the Jacobian are given by $-\mu + \Lambda_1$ and $-\mu + \Lambda_2$, where $\Lambda_1$ and $\Lambda_2$ are the solutions of
\begin{equation}
\Lambda^2 + \left[ \lamL (1-a) + \lamH a - \frac{\lamL \lamH \beta}{\lamL (1-a) + \lamH a} \right]\Lambda
- \lamL \lamH \beta = 0.
\label{eq:Lambda variants}
\end{equation}
The difference in the multiplicative factor 2, which is present in the two terms proportional to $\beta$ in Eq.~\eqref{eq:Lambda} and absent in Eq.~\eqref{eq:Lambda variants}, originates from the fact that the original model, but not the two variants, allows bidirectional infection between an activated node and its neighbor. Therefore, this difference is not essential. The crucial difference between the original model and the two variants are the last term in Eq.~\eqref{eq:Lambda}. This term originates from the simultaneous presence of the two infection processes on the same edge, where an activated infected node can infect a susceptible neighbor and an activated susceptible node can be infected by an infected neighbor.

By substituting Eqs.~(3b), (3d), (4a), and (4b) into Eq.~\eqref{eq:Lambda variants}, one obtains
$\Lambda = \beta/\langle \tau\rangle$, $-1/\langle \tau\rangle$.
Therefore, one obtains $\mu_{\rm c} = \beta/\langle \tau\rangle$, or $\lambda_{\rm eff} \equiv \beta/\langle\tau\rangle \mu_{\rm c} = 1$, which implies that the distribution of inter-event times does not affect the epidemic threshold.

\begin{figure}
\includegraphics[width=8cm]{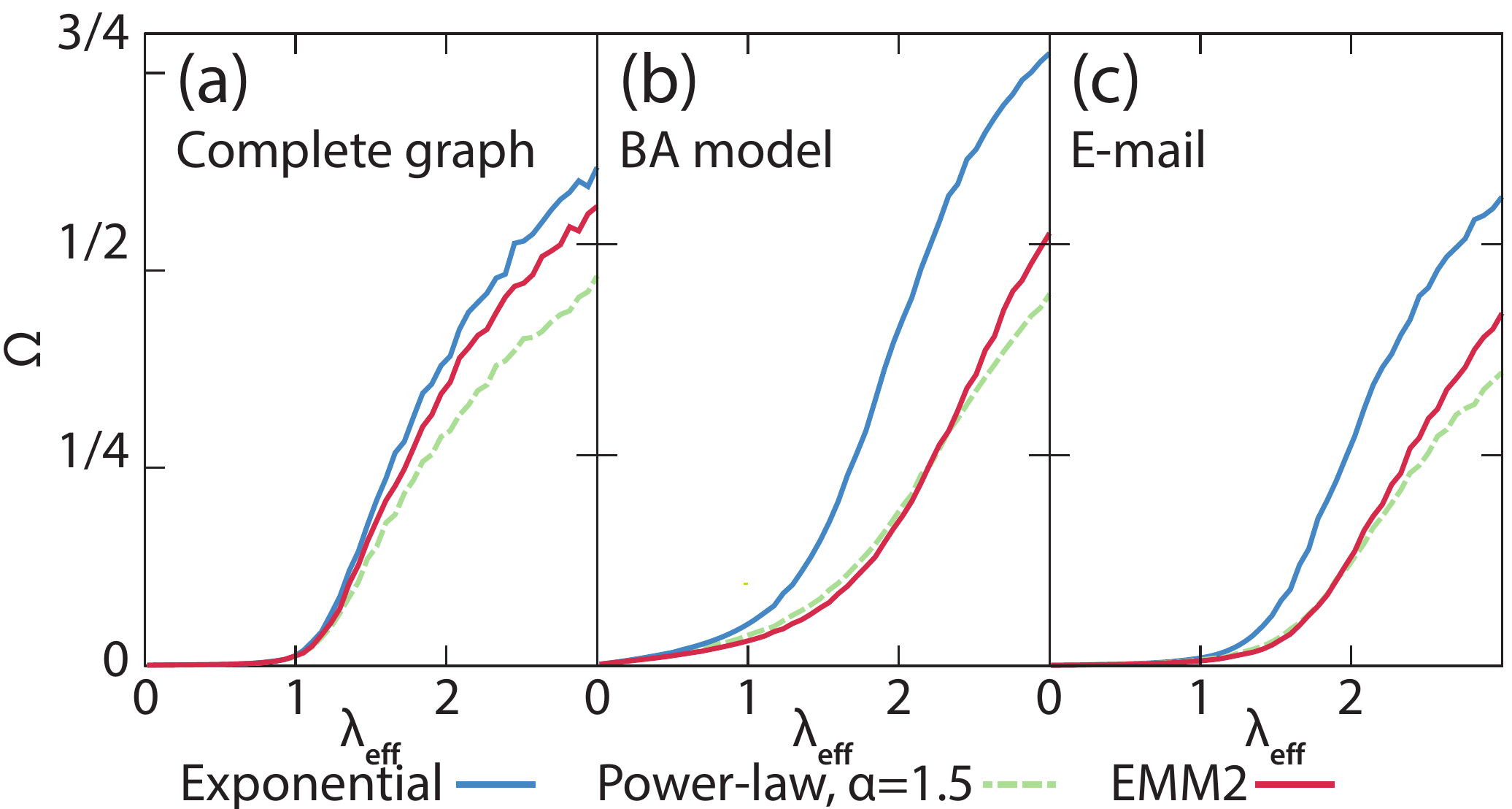}
\caption{Final size of the variant of the node-centric SIR model in which an activated susceptible node can be infected by its infected neighbor, whereas an activated infected node does not infect its susceptible neighbor. (a) Complete graph with $N=10^3$ nodes.
(b) BA model with $N=10^3$ and $m=m_0 = 3$. (c) Email network. We set $\alpha=1.5$ for PL1, and $a=0.3$ and $\text{CV}=\sqrt{5}$ for EMM2.}
\label{fig:nR VM}
\end{figure}

\begin{figure}
\includegraphics[width=8cm]{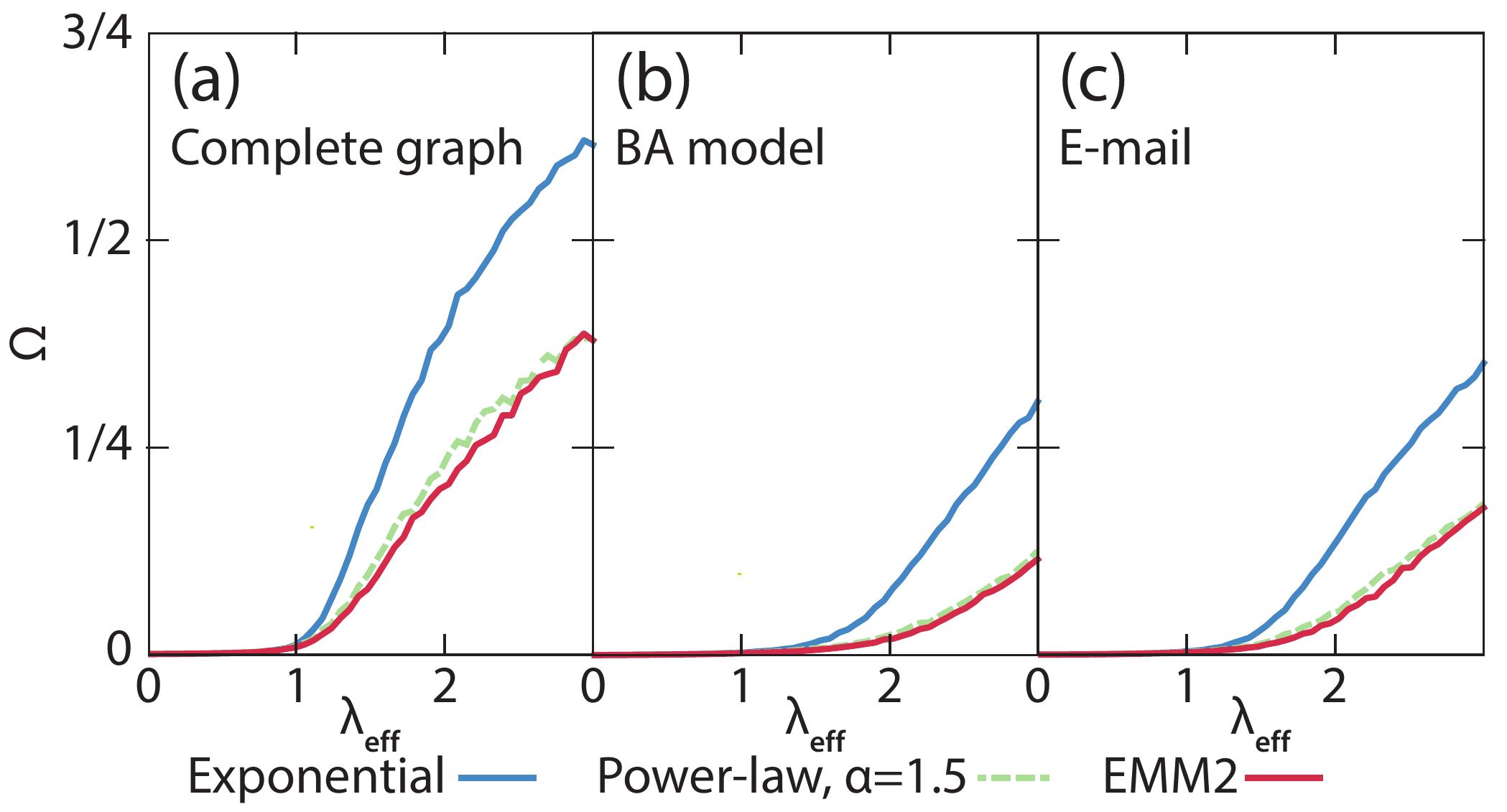}
\caption{Final size of the variant of the node-centric SIR model in which an activated susceptible node is not infected by its infected neighbor, whereas an activated infected node can infect its susceptible neighbor. (a) Complete graph with $N=10^3$ nodes.
(b) BA model with $N=10^3$ and $m=m_0 = 3$. (c) Email network. We set $\alpha=1.5$ for PL1, and $a=0.3$ and $\text{CV}=\sqrt{5}$ for EMM2.
%
}
\label{fig:nR IP}
\end{figure}

In fact, numerically obtained final sizes are smaller for PL1 or EMM2 than for the exponential $\psi(\tau)$ with the same mean on the complete graph and different networks (Figs.~\ref{fig:nR VM} and \ref{fig:nR IP}).

\section{Uniform mixture model having $\langle \tau \rangle=1$ and $\langle \tau^2 \rangle = 2$\label{sec:UMM2 analysis}}

For the UMM2 given by
\begin{equation}
\psi(\tau) = \frac{1-a}{\epsilon} I_{[0, \epsilon]} +
\frac{a}{\epsilon} I_{[\tilde{\tau}, \tilde{\tau} + \epsilon]},
\end{equation}
where $I$ is the indicator function, a straightforward calculation verifies
\begin{equation}
\langle \tau^2 \rangle = \frac{\epsilon^2}{3} + a\tilde{\tau}(\tilde{\tau}+\epsilon).
\label{eq:tau^2 UMM}
\end{equation}
By substituting $\tilde{\tau} = (2-\epsilon)/2a$, which guarantees $\langle \tau\rangle = 1$, into Eq.~\eqref{eq:tau^2 UMM} and imposing $\langle \tau^2\rangle = 2$, one obtains
\begin{equation}
(-2a+3)\epsilon^2 + 12(a-1)\epsilon - 12 (2a-1) = 0.
\label{eq:eps algebraic equation}
\end{equation}
Equation~\eqref{eq:eps algebraic equation} yields
\begin{subequations}
\begin{equation}
\epsilon_1 = \frac{2\left[3(1-a)+\sqrt{3a(2-a)}\right]}{3-2a}
\end{equation}
and 
\begin{equation}
\epsilon_2 = \frac{2\left[3(1-a)-\sqrt{3a(2-a)}\right]}{3-2a}.
\end{equation}
\end{subequations}
It should be noted that the condition $\tilde{\tau} = (2-\epsilon)/2a > 0$ yields
\begin{equation}
\epsilon < \frac{2}{2a+1}.
\label{eq:cnd on epsilon}
\end{equation}

In fact, $\epsilon = \epsilon_1$ violates Eq.~\eqref{eq:cnd on epsilon} because $\epsilon_1 < 2/(2a+1)$ yields $(12a^2+1)(2a-3) > 0$, which is a contradiction given $0<a<1$. In contrast, $\epsilon = \epsilon_2$ satisfies Eq.~\eqref{eq:cnd on epsilon} and $\epsilon_2 > 0$. Therefore, $\epsilon = \epsilon_2$ is the unique solution.

\bibliographystyle{abbrv}
\bibliography{citations}

\end{document}